\documentclass[apj]{emulateapj}

\newcommand{\etal}{et al.}

\newcommand\rosat{{\it ROSAT\/}}

\newcommand\xte{{\it RXTE\/}}

\newcommand\chandra{{\it Chandra}}

\newcommand\xmm{{\it XMM-Newton}}

\newcommand\hess{{HESS\/}}

\newcommand\suzaku{{\it Suzaku\/}}
\newcommand\swift{{\it Swift\/}}
\newcommand\cco{CXOU~J171405.7$-$381031}

\newcommand\ctb{CTB~37B}
\newcommand\tevtwo{HESS~J1713$-$381}

\newcommand\snr{G353.6$-$0.7}
\newcommand\tev{HESS~J1731$-$347}
\newcommand\psr{XMMU~J173203.3$-$344518}

\def\simlt{\mathrel{\hbox{\rlap{\hbox{\lower4pt\hbox{$\sim$}}}\hbox{$<$}}}}
\def\simgt{\mathrel{\hbox{\rlap{\hbox{\lower4pt\hbox{$\sim$}}}\hbox{$>$}}}}

\slugcomment{}

\shorttitle{Two Magnetar Candidates}
\shortauthors{Halpern \& Gotthelf}

\begin{document}

\title{Two Magnetar Candidates in HESS Supernova Remnants}

\author{J. P. Halpern and E. V. Gotthelf}
\affil{Columbia Astrophysics Laboratory, Columbia University,
550 West 120th Street, New York, NY 10027-6601, USA;
jules@astro.columbia.edu, eric@astro.columbia.edu}

\begin{abstract}
We identify two candidate magnetars in archival X-ray observations
of \hess\ detected shell-type SNRs.  X-ray point sources in \ctb\ coincident
with \tevtwo\ and in \snr\ coincident with \tev\ both have AXP-like spectra,
much softer than those of ordinary, rotation powered pulsars, and no
optical/IR counterparts.  The spectrum of \cco\ in \ctb\ has a hard
excess above 6~keV, which may be similar to such components seen in some
AXPs.  A new \chandra\ observation of this object reveals
a highly significant pulsed signal at $P = 3.82$~s with
pulsed fraction $f_p = 0.31$.
Analysis of an \xmm\ observation of the second candidate,
\psr\ in \snr, yields only marginal
evidence for a 1~s period.  If it is not a magnetar, then
it could be a weakly magnetized central compact object (CCO).
Considering
that these \hess\ sources previously attributed to the SNR shells
are possibly
centrally peaked, we hypothesize that their pulsars may contribute
to diffuse TeV emission.  These identifications potentially
double the number of magnetar/SNR associations in the Galaxy,
and can be used to investigate the energetics and asymmetries
of the supernovae that give rise to magnetars.
\end{abstract}

\keywords{ISM: individual (\ctb, \snr ) ---
pulsars: individual (\cco, \psr) --- stars: neutron }

\section {Introduction}

Surveys by the atmospheric Cherenkov telescopes HEGRA, \hess,
MAGIC, and VERITAS, have revealed more than 50 Galactic TeV
sources\footnote{VHE $\gamma$-ray Sky Map and Source Catalog,
http://www.mppmu.mpg.de/\~\,rwagner/sources/index.html},
the majority of which are supernova remnants (SNRs) or pulsar wind
nebulae (PWNe), the latter being the largest class.
For reviews of TeV PWNe, see \citet{car08}, \citet{gal08},
\citet{lem08}, \citet{hes08}, and \citet{mat09}.
In almost all cases, the pulsars responsible for the PWNe that
are detected at TeV energies have spin-down luminosities
$> 10^{36}$ erg~s$^{-1}$.  Exceptions are the
very nearby Geminga pulsar ($\dot E = 3.2 \times 10^{34}$
erg~s$^{-1}$), with diffuse TeV emission detected by Milagro
\citep{abd09}, and PSR J2032+4127 ($\dot E = 2.7 \times 10^{35}$
erg~s$^{-1}$) associated with TeV J2032+4130 in the Cyg OB2
stellar association \citep{cam09}.
There is no evidence that any of the 14 previously
known magnetars\footnote{SGR/AXP Online Catalog,
http://www.physics.mcgill.ca/\~\,pulsar/magnetar/main.html}
produce TeV emission.  Their spin-down luminosities,
$\dot E \equiv 4\pi^2I\dot P/P^3$,
are $< 10^{35}$ erg~s$^{-1}$, less than that of known pulsars
producing X-ray PWNe, and it is not even certain that 
they accelerate particles to TeV energies as do
ordinary radio and spin-powered pulsars.

\citet{aha06} discovered the TeV source \tevtwo\
and associated it with the SNR \ctb.  Using a \chandra\
observation, \citet{aha08a} then found the point X-ray source
\cco\ in \ctb, and considered that it could be a pulsar,
albeit with an unusually soft, non-thermal spectrum.
\citet{nak09} analyzed \chandra\ and \suzaku\ spectra of
\cco, noting also possible variability in flux,
which they took to be good evidence that it is an
anomalous X-ray pulsar (AXP).  In Section 2 of this paper,
we report the discovery of pulsations from \cco\ that
verifies this conjecture.  We also
analyze archival X-ray data on another compact source
in a \hess\ detected SNR, \snr=\tev\
\citep{tia08,aha08b}, showing that it too is a magnetar
candidate based on its spectrum and a possible periodicity.
Considering these new associations, we discuss in
Section 3 the possibility that an early stage of a magnetar
may produce a TeV PWN, a
precedent being PSR~J1846$-$025 in Kes~75 \citep{dja08}.
We conclude in Section 4 that it is of great interest to measure
the spin-down rates of these new candidate magnetars to
test this hypothesis.

\section {X-ray Observations of New Candidate Magnetars}

\subsection{\tevtwo/\ctb}

At radio wavelengths, \ctb\ is a shell $\approx 10^{\prime}$ in diameter
\citep{kas91}.  Its distance, estimated from \ion{H}{1} absorption
and the Galactic rotation curve, is $\approx 8$~kpc \citep{gre06}.
Its age may be only $\sim 1500$~yr, which may associate it with the
historical supernova 373~AD \citep{cla75}.
Following the discovery of \tevtwo\ by \citet{aha06},
further \hess\ observations by \citet{aha08a}
localized it to the center of the radio shell of \ctb, and
found that the extent of the TeV source is compatible with
either a centrally peaked Gaussian of 
$\sigma = 2.\!^{\prime}6 \pm 0.\!^{\prime}8$
or a shell of radius $4^{\prime}$--$6^{\prime}$,
consistent with the size of the radio remnant.
The energy flux of \tevtwo\ from 0.2--10~TeV
is $\approx 4.2 \times 10^{-12}$ erg~cm$^{-2}$~s$^{-1}$,
corresponding to luminosity
$L_{\gamma} = 3.2 \times 10^{34}\ d_8^2$ erg~s$^{-1}$,
where $d_8$ is the distance in units of 8~kpc.

%-----------------------------Figure Start------------------------------
\begin{figure*}
\centerline{
\hfill
\includegraphics[scale=0.52,angle=270]{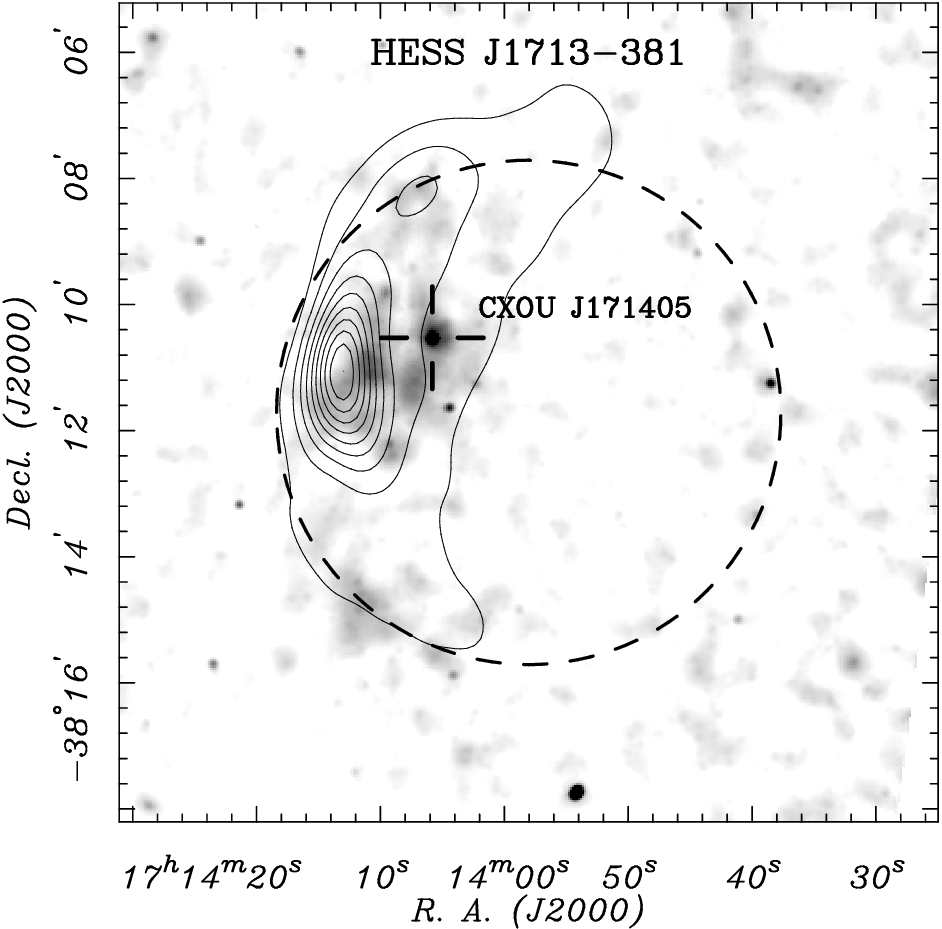}
\hfill
\includegraphics[scale=0.52,angle=270]{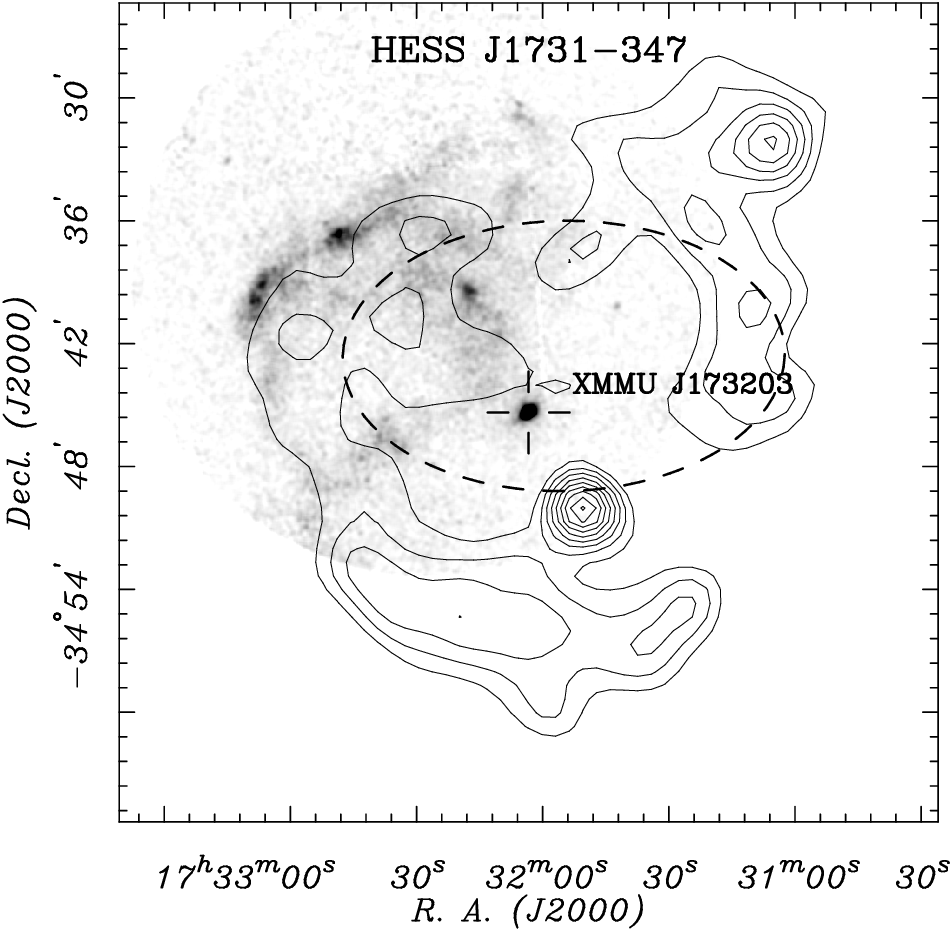}
\hfill
}
\caption{\hess\ TeV sources denoted by dashed ellipses
representing their estimated intrinsic extent (Aharonian et al. 2008a,
2008b).  X-ray point sources (crosses) are magnetar candidates.
Left: \tevtwo\ with the
shell-type radio SNR \ctb\ (contours); MOST 843~MHz) and X-ray
emission (greyscale; \chandra\ ACIS-I).
Right: \tev\ associated with the shell-type radio SNR
\snr\ (contours; ATCA 843~MHz) and its X-ray emission
(greyscale; \xmm\ EPIC MOS).  The bright, compact
radio source SSW of the pulsar is extragalactic, based on \ion{H}{1}
absorption velocities (Tian et al. 2008).  The radio source to the
NW is an \ion{H}{2} region at 3.2~kpc, which is presumed to be associated
with the SNR.
}
\label{fig1}
\end{figure*}
%-----------------------------Figure End--------------------------------

\begin{deluxetable*}{lllclrc}
\tablewidth{0pt}
\tablecaption{Log of X-ray Observations of HESS Sources}
\tablehead{
& & \colhead{Instrument/} & \colhead{ObsID/} & \colhead{Date} & \colhead{Exp.} & \\
\colhead{Source} & \colhead{Mission} \hfill & \colhead{Mode} & \colhead{Seq. No.}
& \colhead{(UT)} & \colhead{(s)} & \colhead{Refs.}
}
\startdata
J1713$-$381  &  \suzaku\   &   XIS/Normal       &   501007010  &  2006 Aug 27  & 82,815  &  1      \\
&  \chandra\  &   ACIS-I/TE/VF     &        6692  &  2007 Feb 2   & 25,155  &  1,2,3  \\
&  \xte\      &   PCA            & 93011-03-01  &  2008 Jul 28-31 & 89,326  &  3      \\
&  \chandra\  &   ACIS-S/CC        &       10113  &  2009 Jan 25  & 30,146  &  3      \\
\hline
J1731$-$347  &  \suzaku\   &   XIS/Normal       &   401099010  &  2007 Feb 23  & 40,619  &  3,4    \\
&  \xmm\      &   pn+MOS/FF        &  0405680201  &  2007 Mar 21  & 25,408  &  3,4,5  \\
&  \chandra\  &   ACIS-I/TE/VF     &        9139  &  2008 Apr 28  & 29,610  &  3,5    \\  
&  \swift\    &   XRT/PC           & 00037740001  &  2009 Feb 4   &  1,387  &  3      \\  
&  \swift\    &   XRT/PC           & 00037740002  &  2009 Mar 15  &  1,432  &  3 
\enddata
%\tablecomments{}
%\tablenotetext{a}{\footnotesize}
\tablerefs{(1) Nakamura \etal\ 2008; (2) Aharonian \etal\ 2008a;  (3) This work;
(4) Tian \etal\ 2009; (5)  Acero \etal\ 2009.}
\label{logtable}
\end{deluxetable*}

The \chandra\ ACIS-I observation listed in Table~\ref{logtable}
revealed a bright X-ray point source and
faint, diffuse emission that \citet{aha08a} fitted to
a non-equilibrium ionization model.
In Figure~\ref{fig1} we show the \chandra\ image with
superposed radio contours from the Molonglo Galactic
Plane Survey at 843~MHz \citep{gre99}, and the $1\sigma$
extent of the TeV emission.  Although only the eastern
edge of the radio SNR is seen here, at lower frequency a
nearly circular shell is evident \citep{kas91}.
The point source \cco\ lacks an optical or IR counterpart
in the Digitized Sky Survey or in 2MASS, and is
likely associated with the SNR based on its location, flux, and
spectrum.  \citet{aha08a} noted that an absorbed power-law fit
to the point source,
with photon index $\Gamma = 3.3^{+0.2}_{-0.1}$, yields an
$N_{\rm H}$ that is consistent with the fit to the diffuse emission.
Alternatively, they obtained a good fit with a two-temperature
blackbody of $kT_1 = 0.52$~keV and $kT_2 = 1.6$~keV.
There is no strong evidence for a PWN associated with the point source.

Using the \suzaku\ observation listed in Table~\ref{logtable},
\citet{nak09} found nonthermal X-ray emission in the
southern region of \ctb.  They also reanalyzed the \chandra\
spectrum of the point source, finding a 2--10~keV luminosity
of $\approx 1.4 \times 10^{34}\ d_8^2$ ergs~s$^{-1}$.  Comparing
that with the \suzaku\ luminosity, which appeared to be a factor
of 1.8 higher, they conjectured that variability plus the
steep X-ray spectrum indicated that \cco\ is an AXP.

\begin{figure}[t]
\centerline{\includegraphics[scale=0.35,angle=270]{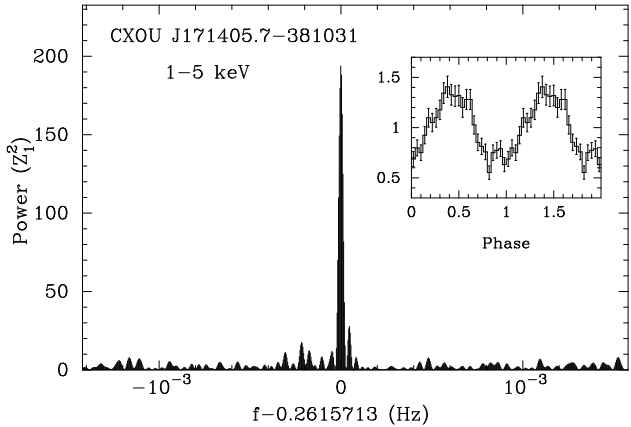}}
\caption{$Z_1^2$ power spectrum of \cco\ from the \chandra\ CC-mode
observation, and folded light curve (inset) of the new 3.82~s pulsar.
Normalized count rate is plotted.
%({\it b}) $Z_1^2$ power spectrum from the \xmm\ observation of \psr.
%Only a tentative detection of 1.01~s pulsations can be suggested
%(see \S 2.2).
}
\label{fig2}
\end{figure}

We obtained a new \chandra\ observation of \cco\ on 2009 January 25
using ACIS-S in continuous clocking (CC) mode for 30~ks
to search for the expected pulsations.
After barycentric correction, we searched for a periodic signal
to the Nyquist limit of 5.7~ms.  We find a highly significant
signal at a period of $3.82305 \pm 0.00002$~s, having a single broad
pulse with pulsed fraction $f_p = 0.31$, defined as the fraction
of counts above the minimum in the light curve.
Figure~\ref{fig2} shows the
folded light curve in the 1--5~keV band, which contains most of the
counts.  The peak power measured using the $Z_1^2$ statistic is 194,
corresponding to a vanishing chance probability.
There is no evidence for energy dependence of the pulse profile.
The long period is consistent with expectation based on the
AXP-like spectrum, and falls in the 2--12~s range seen for magnetars.
It is possible that the true period is 7.6~s,
with two peaks per rotation, but there is no strong evidence to
support this in the form of distinguishable peaks in the current data.

Listed in Table~\ref{logtable} is a set of archival \xte\ 
observations that were pointed
$14^{\prime}$ from the position of \cco, well within
the response of the Proportional Counter Array.
We searched these data for the 3.82~s period,
but no significant signal was found.  Given the
\chandra\ measured spectrum and flux of the source (discussed next),
the lack of detection by \xte\ is not surprising.

%-----------------------------Figure Start------------------------------
%-----------------------------Figure End--------------------------------

\begin{deluxetable}{lcc}
\tablewidth{0pt}
\tablecaption{Spectral Fits to \chandra\ ObsID 6692: \cco}
\tablehead{
\colhead{Parameter}  & \colhead{PL} & \colhead{PL + BB}
}
\startdata
$N_{\rm H}$~($10^{22}$ cm$^{-2}$)        & $4.03^{+0.51}_{-0.47}$  & $2.59\pm0.43$ \\
$\Gamma$                                 & $3.10\pm0.32$           & $-2.0^{+1.1}_{-1.0}$   \\
$kT$ (keV)                      & . . .                   & $0.64^{+0.07}_{-0.04}$ \\
$R$ (km)                        & . . .                   & $0.82\,d_8$ \\ 
$F_x(2-10\ {\rm keV}$)\tablenotemark{a}  & $1.8 \times 10^{-12}$   & $1.8 \times 10^{-12}$  \\
$L_x(2-10\ {\rm keV}$)\tablenotemark{b}  & $1.4 \times 10^{34}\,d_8^2$    & $1.7 \times 10^{34}\,d_8^2$   \\
$\chi^2_{\nu}(\nu)$                      & 1.28(96)                & 1.06(94)
\enddata
\tablenotetext{a}{Absorbed flux in units of erg cm$^{-2}$ s$^{-1}$.}
\tablenotetext{b}{Unabsorbed luminosity in units of erg s$^{-1}$.}
\label{spec1table}
\end{deluxetable}

\begin{figure}[t]
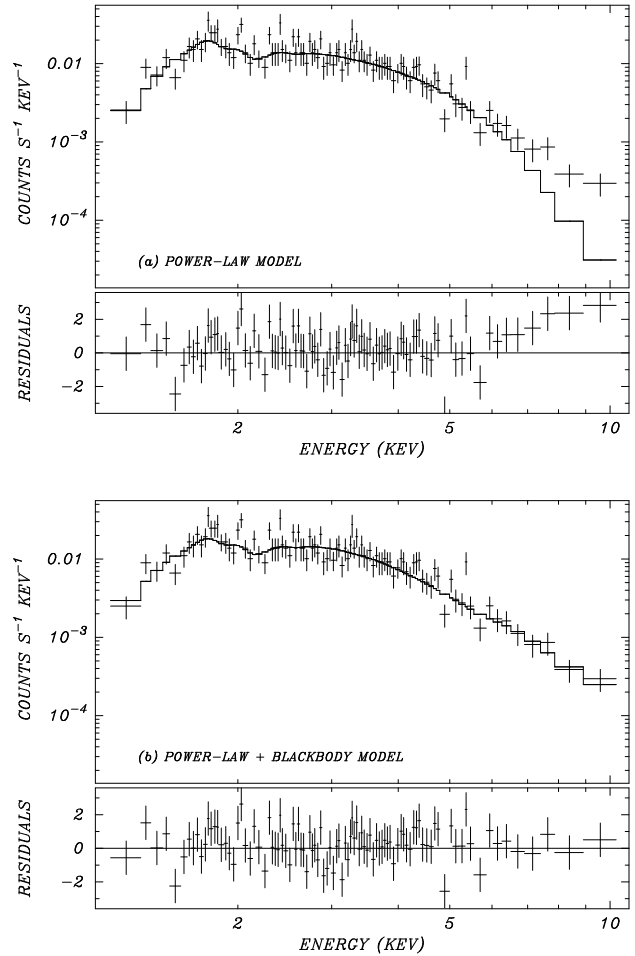

\centerline{\includegraphics[scale=0.35,angle=270]{f3a.eps}}
\vspace{0.2in}
\centerline{\includegraphics[scale=0.35,angle=270]{f3b.eps}}
\caption{({\it a}) Spectrum of \cco\ from t e \chandra\ ACIS-I/TE
observation fitted to a power-law model.  Positive residuals are
evident above 6 keV.
({\it b}) The same spectrum fitted to a power-law plus blackbody model
(see Section 2.1).  Fitted parameters are listed in Table~\ref{spec1table}.}
\label{fig3}
\end{figure}

We reanalyzed the spectrum from the ACIS-I observation,
fitting a power-law or blackbody model as before.  The
results are listed in Table~\ref{spec1table}.  For the power-law
model, we obtain a photon index $\Gamma = 3.1$, consistent
with the published analyses.  However, this fit is not
acceptable (see Figure~\ref{fig3}), as there is clear excess
of flux above 6~keV. This deviation is not seen in \citet{aha08a}
and \citet{nak09}, evidently because they excluded channels
above 7~keV.  A single blackbody is even a worse fit,
and a two blackbody model requires an unphysical high temperature.
Although we can get a adequate fit using a power-law plus
blackbody model, the resulting negative spectral index,
$\Gamma = -2.0^{+1.1}_{-1.0}$ is highly unusual, and
possibly not meaningful.  This hard X-ray
excess may be an extreme case of the components that dominate
at energies above 10~keV in several AXPs
\citep{mol04,kui04,rev04,den06}, although the latter
have spectral indices $0.9 < \Gamma < 1.4$.
A more unusual possibility would be a broad cyclotron feature 
overlapping the end of the spectrum.
The fitted blackbody component, however,
with $kT_{\rm BB} = 0.64^{+0.07}_{-0.04}$~keV, is typical
for an AXP.   It has a radius of $0.8\,d_8$~km, and a
bolometric luminosity of 
$1.4 \times 10^{34}\ d_8^2$ ergs~s$^{-1}$.

%-----------------------------Figure Start------------------------------
%-----------------------------Figure End--------------------------------

The flux from the second \chandra\ observation of this
source is similar to that of the first; however, we cannot use
it to address the peculiarities of the spectrum because
of the greatly enhanced background in CC-mode.

\subsection{\tev/\snr}

\tev\ is an extended source that coincides with a patch of diffuse
X-ray emission in the \rosat\ All-Sky Survey \citep{aha08b}.
\cite{tia08} discovered that the X-ray feature is associated
with multiwavelength radio emission that comprises a new SNR, \snr.
A two-dimensional Gaussian fit to the TeV source gives semi-major
and semi-minor axes of $\sigma_1 = 0.\!^{\circ}18 \pm 0.\!^{\circ}07$
and $\sigma_2 = 0.\!^{\circ}11 \pm 0.\!^{\circ}03$ for the main part
of the emission, after allowing for the instrumental resolution
\citep{aha08b}.  The TeV source thus fits within the $0.\!^{\circ}5$ diameter
radio shell (Figure~\ref{fig1}), while the diffuse \rosat\ emission
is a half shell matching the southern Galactic half of the radio
remnant \citep[see][]{tia08}.
The other half is likely absorbed in soft X-rays by the
increasing molecular gas density toward the Galactic plane.

\citet{tia08} assumed that the SNR is associated with an
an adjacent \ion{H}{2} region at $d=3.2$~kpc. \citet{tia09}
found a molecular cloud at the same velocity as an \ion{H}{1}
absorption feature that they consider to be associated with
the \ion{H}{2} region.
Considering the implied $\approx 28$~pc diameter of the SNR,
they suggested that \snr\ is $\sim 27,000$ years old and has
entered the radiative phase.  This, they argued, is the best
evidence that an old SNR can still accelerate particles to
greater than TeV
energies, as predicted by \citet{yam06}, and that the TeV
emission comes from the interaction of the SNR shock with
the molecular cloud.
However, while the outline of \tev\ follows the radio and
X-ray contours, it is not entirely clear whether the
the TeV emission is peaked in the center or follows the
shell.  Some evidence for the latter is seen in newer
HESS data \citep{ace09}.   There is also a ``tail'' of
TeV emission to the west, beyond the radio shell, which may
or may not be part of the same source \citep{aha08b}.
The energy flux of \tev\ above 0.5~TeV
is $\approx 3.3 \times 10^{-11}$ erg~cm$^{-2}$~s$^{-1}$,
corresponding to luminosity
$L_{\gamma} = 4.0 \times 10^{34}\ d_{3.2}^2$ erg~s$^{-1}$.

\suzaku, \xmm, \chandra, and \swift\ observations of \tev\
are listed in Table~\ref{logtable}.  Figure~\ref{fig1} shows
the \xmm\ image, which reveals the filaments of \snr, as well as
a point source near the center of the radio shell that was not
detected in the \rosat\ All-Sky Survey.
We began by fitting the spectrum of a region of bright, diffuse SNR emission
in the \suzaku\ data.  Photons from an elliptical region
$\approx 6^{\prime}$ long, adjacent to the point source and extending to
the northeast of it, were extracted from all three XIS
detectors and fitted to an absorbed power-law spectrum.  The
resulting photon index is $\Gamma = 2.54 \pm 0.15$, with an
absorbing column density 
$N_{\rm H} = (1.65 \pm 0.17) \times 10^{22}$~cm$^{-2}$.  There
is no evidence for emission lines in the spectral residuals.
More complete analysis of the spatially resolved spectra
\citep{ace09,tia09} reveal ubiquitous non-thermal emission, with
a gradient of $N_{\rm H}$ across the remnant
that is consistent with additional intervening material toward the
Galactic plane.
The spectra thus indicates the essentially
nonthermal nature of the SNR, similar to the brightest
TeV SNR RX~J1713.7$-$3946.

We next turned to the compact source \psr.  It lies off-axis
in both the \xmm\ and \chandra\ images.  The best position, from the
\chandra\ image, is R.A. = $17^{\rm h}32^{\rm m}03.\!^{\rm s}41$,
decl = $-34^{\circ}45^{\prime}16.\!^{\prime\prime}6$ (J2000.0).
It lacks an optical or IR counterpart in the Digitized
Sky Survey or in 2MASS, and is therefore a candidate for a
neutron star in \snr.

\begin{deluxetable}{lcc}
\tablewidth{0pt}
\tablecaption{Spectral Fits to \psr }
\tablehead{
\colhead{Parameter}  & \colhead{\suzaku} & \colhead{\xmm}
}
\startdata
& Power-law Model & \\
\colrule
$N_{\rm H}$~($10^{22}$ cm$^{-2}$)        & $3.3\pm0.2$  & $3.2\pm0.1$  \\
$\Gamma$                                 & $4.5\pm0.2$  & $4.7\pm0.1$  \\
$F_x(0.5-10\ {\rm keV}$)\tablenotemark{a}& $3.5 \times 10^{-12}$  & $2.8 \times 10^{-12}$  \\
$L_x(0.5-10\ {\rm keV}$)\tablenotemark{b}& $1.8 \times 10^{35}\,d_{3.2}^2$  & $2.0 \times 10^{35}\,d_{3.2}^2$ \\
$\chi^2_{\nu}(\nu)$                      & 0.98(136)    & 1.87(148) \\
\colrule
& Blackbody Model & \\
\colrule
$N_{\rm H}$~($10^{22}$ cm$^{-2}$)        & $1.52\pm0.09$  & $1.40\pm0.06$ \\
$kT$ (keV)                      & $0.50\pm0.01$  & $0.49\pm0.01$  \\
$R$ (km)                        & $1.1\,d_{3.2}$ & $1.0\,d_{3.2}$ \\ 
$F_x(0.5-10\ {\rm keV}$)\tablenotemark{a}& $3.2 \times 10^{-12}$   & $2.8 \times 10^{-12}$ \\
$L_x(0.5-10\ {\rm keV}$)\tablenotemark{b}& $9.1 \times 10^{33}\,d_{3.2}^2$  & $7.7 \times 10^{33}\,d_{3.2}^2$\\
$\chi^2_{\nu}(\nu)$                      & 1.0(136)                & 1.15(148) \\
\colrule
& Two Blackbody Model & \\
\colrule
$N_{\rm H}$~($10^{22}$ cm$^{-2}$)        & $1.80^{+0.27}_{-0.18}$  & $1.66^{+0.23}_{-0.12}$  \\
$kT_{1}$ (keV)                    & $0.42^{+0.07}_{-0.05}$  & $0.40^{+0.03}_{-0.08}$  \\
$R_{1}$ (km)                      & $1.6\,d_{3.2}$   & $1.5\,d_{3.2}$   \\ 
$kT_{2}$ (keV)                    & $0.85^{+0.25}_{-0.16}$  & $0.64\pm0.10$   \\
$R_{2}$ (km)                      & $0.17\,d_{3.2}$   & $0.36\,d_{3.2}$  \\ 
$F_x(0.5-10\ {\rm keV}$)\tablenotemark{a}& $3.4 \times 10^{-12}$   & $2.8 \times 10^{-12}$  \\
$L_x(0.5-10\ {\rm keV}$)\tablenotemark{b}& $1.1 \times 10^{34}\,d_{3.2}^2$  & $9.4 \times 10^{33}\,d_{3.2}^2$ \\
$\chi^2_{\nu}(\nu)$                      & 0.81(134)                & 0.99(145)
\enddata
\tablenotetext{a}{Absorbed flux in units of erg cm$^{-2}$ s$^{-1}$, average of all detectors.}
\tablenotetext{b}{Unabsorbed luminosity in units of erg s$^{-1}$.}
\label{spec2table}
\end{deluxetable}

\begin{figure}
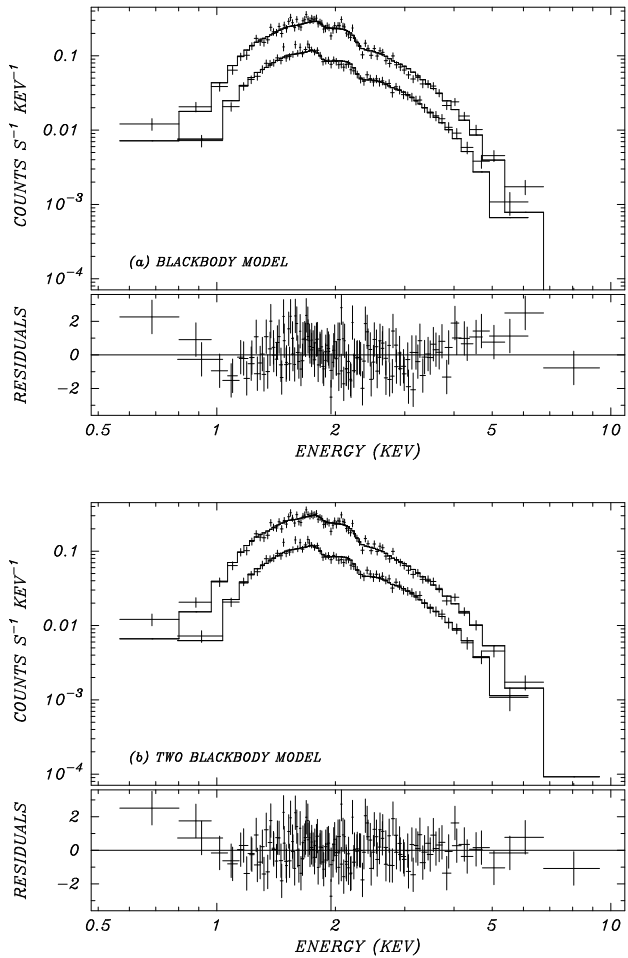

\centerline{\includegraphics[scale=0.35,angle=270]{f4a.eps}}
\vspace{0.2in}
\centerline{\includegraphics[scale=0.35,angle=270]{f4b.eps}}
\caption{({\it a}) Spectrum of \psr\ from the \xmm\
observation, fitted to a blackbody model.  The upper spectrum is
from the pn CCD, and the lower spectrum is from the two MOS detectors.
({\it b}) The same spectra fitted to a two-blackbody model
(see Section 2.2).  Fitted parameters are listed in Table~\ref{spec2table}.}
\label{fig4}
\end{figure}

Power-law spectral fits to the \suzaku\ and \xmm\
data for \psr\ have very steep slopes,
$\Gamma = 4.5-5.7$ as listed in Table~\ref{spec2table}, 
and the power-law fit to the \xmm\ data is poor.  The \chandra\
spectrum is not as steep ($\Gamma = 3.98\pm0.13$), but
it should be disregarded. As a strong source
with count rate 0.21~s$^{-1}$, it is affected by pileup,
which tends to flatten the spectrum.
A better and more physically motivated
fit to the \suzaku\ and \xmm\ data
is obtained with a blackbody of
$kT \approx 0.5$ and radius of 1~km.  This is typical
for a CCO, but also for an AXP.
A good fit is also obtained using two blackbodies,
with $kT_1 \approx 0.4$~keV and $kT_2$
in the range 0.6--0.9~keV.  The single and
double blackbody fits are shown in Figure~\ref{fig4}.
The higher temperature component is also seen in AXPs, but is
somewhat higher than has been seen in any CCO,
which tend to cluster around 0.4~keV
\citep{kar02,laz03,del04,hal07,got09,pav09}.
Adopting the two-blackbody model, we obtain a bolometric luminosity of
$\approx 1 \times 10^{34}\ d_{3.2}^2$ erg~s$^{-1}$,
which is in the range of quiescent AXPs, but on the high side for
a CCO \citep{got08}.
The fitted column densities in the thermal models
of the compact source
are consistent with that of the diffuse
nonthermal SNR emission, which favors these models
and lends further support to the association 
of the compact source with the SNR.

\begin{figure}
\vspace{-0.26in}
\centerline{\includegraphics[scale=0.35,angle=270]{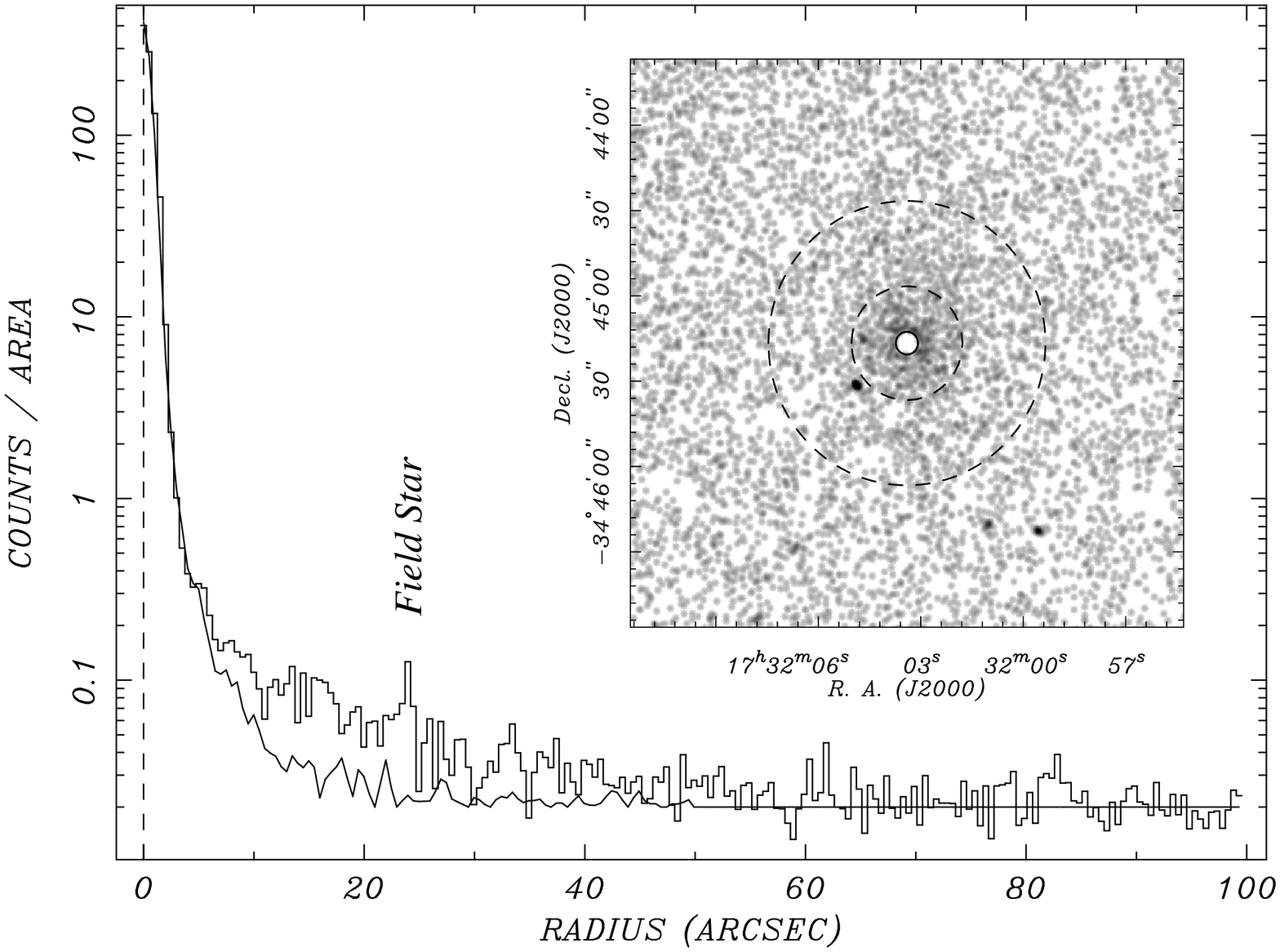}}
\vspace{-0.26in}
\caption{Nebulosity associated with \psr\ in the \chandra\ image.
The radial profile (histogram) is compared to a
{\tt CHART/MARX} simulation of a point source (solid curve) using
the position and spectrum of \psr, and allowing for CCD pileup.
``Field star'' refers to the point source to the southeast.
Inset: The \chandra\ image in which the compact component of \psr\ has been
excised (solid circle).  The dahed circles, of radii $20^{\prime\prime}$
and $50^{\prime\prime}$, enclose the regions used for the spectral fits
and flux calculations discussed in Section 2.2.
}
\label{fig5}
\end{figure}

The \chandra\ image shows faint, symmetric nebulosity
around \psr. Excess emission compared to the point spread
function, modelled at the off-axis location of the piled-up
point source,
is seen out to $\sim 50^{\prime\prime}$ (Figure~\ref{fig5}).
It is not immediately obvious whether this is a PWN or a
dust scattered halo, although its symmetry suggests the
latter.   We extracted counts from an annulus
$5.5^{\prime\prime} < r < 50^{\prime\prime}$, and
fitted its spectrum to a power-law model with column density
fixed at the same value as for a power-law fit to the point source.
This yields an acceptable
fit with $\Gamma = 4.97 \pm 0.14$ and a flux that is only 9\%
of that of the point source.  The measured spectrum
is anomalously soft for a PWN, but is slightly softer
than that of the point source, which is expected for a
dust scattered halo.  To further check this hypothesis,
we fitted the spectrum from a smaller annulus,
$5.5^{\prime\prime} < r < 20^{\prime\prime}$ containing
the brightest part of the halo,
and find a spectrum that is consistent with
that of the point source.  In any case,
the absence of a brighter PWN
probably limits the spin-down power of the putative
pulsar to $<4\times 10^{36}$ erg~s$^{-1}$ \citep{got02}.

%-----------------------------Figure Start------------------------------
%-----------------------------Figure End--------------------------------

At the current stage of the analysis there is no
strong evidence for variability of the point source.
The \suzaku, \xmm, and \chandra\ observations
agree to within 20\% in flux.
Two short observations with the \swift\ XRT
listed in Table~\ref{logtable} and
separated by 39 days are useful to search for
X-ray variability that would be a signature
of a magnetar.  The 0.5--10~keV counts
from \psr\ are marginally consistent between
these two observations, with count rates
of $0.056 \pm 0.006$~s$^{-1}$ and
$0.073 \pm 0.007$~s$^{-1}$, respectively.
The significance of the difference is $<2\,\sigma$.
These rates are consistent with the fluxes
measured previously for the source by \suzaku\
and \xmm.  Therefore, no evidence of
magnetar-like activity has been seen over the 2 year
time span during which \psr\ has been observed.

%-----------------------------Figure Start------------------------------
%-----------------------------Figure End--------------------------------

During the \xmm\ observation, the
EPIC pn CCD was operated in full frame mode with a time resolution
of 73.4~ms.  This allows us to search for periods $>150$~ms.
After barycentric correction, we
find only marginal evidence for a period of 1.010441(6)~s from
\psr, modulated with a $\sim 10\%$ pulsed fraction.
%(Figure~\ref{fig2}).
For this signal, $Z_1^2 = 32.7$, corresponding to a 0.4\% chance
probability for a blind search for $P>1$~s, or 2.7\% chance
probability for a search of all independent periods $> 150$~ms.
We consider this too weak to claim a secure detection, although
such a period is plausible in either the AXP or the CCO
scenario, the implications of which are discussed below.

\section{Discussion}

Previous authors have argued that particles accelerated at the
SNR shock are responsible for the TeV emission via pion decay
from both \tevtwo\ \citep{aha08a}
and \tev\ \citep{tia08}.  In the latter case,
interaction with nearby molecular clouds is envisioned
\citep{tia08,tia09}.  On the other hand, neither
TeV source clearly has a pure shell-like morphology,
so it cannot be ruled out that their associated pulsars
make some contribution to the TeV flux via the
inverse Compton mechanism.

Magnetars are not expected to power TeV PWNe because
their large magnetic fields have already spun them down to
where their {\it present\/} spin-down power $\dot E$
is in the range $10^{32}-10^{35}$
erg~s$^{-1}$.  This is much less than the $\dot E = 10^{36-38}$
erg~s$^{-1}$ of pulsars powering TeV sources and PWNe in general.
Also, unlike ordinary pulsars, it is not certain that magnetars can accelerate 
particles to TeV energies.  Magnetar models that involve strong currents
on closed, twisted magnetic field lines develop voltages of only
$\sim 10^9$~V \citep{bel07}.  On the other hand, it is not excluded
that the ordinary pulsar mechanism operates on open magnetic field lines
of magnetars, which may develop particle-dominated winds that
flow out to become shocked PWNe.
Evidence for this possibility comes from the two magnetars
that are also transient radio pulsars \citep{cam06,cam07}.

Even though no TeV source has yet been associated with a known
magnetar, we consider it possible that a magnetar could energize
a PWN at a younger stage, when its period is short and
its spin-down power is large.  Then, some high-energy electrons
may escape the high $B$-field region and emit via inverse Compton
scattering into the TeV band for an extended time.
A possible prototype of such systems is PSR~J1846$-$0258, the
``transitional'' $\sim 700$~yr old pulsar \citep{got00}
associated with HESS~J1846$-$029 in the shell-like SNR Kes 75
\citep{dja08}. This $0.32$~s pulsar has a dipole field that
approaches magnetar strength ($B_s = 4.9 \times 10^{13}$~G) and
displays AXP-like bursts (Gavriil et al. 2008); it is therefore likely
of an intermediate class connecting the rotation-powered and the
magnetar pulsars.  It is also one of the most energetic pulsars;
its $\dot E =  8.1 \times 10^{36}$ erg~s$^{-1}$ is
easily sufficient to power its X-ray PWN and the TeV emission of
HESS~J1846$-$029.

The quantitative details of any magnetar source
model for extended TeV emission can be constrained by
independent age estimates from associated SNRs.
The number of magnetars securely associated
with shell SNRs is arguably 
at most three, 1E~2259+586 in CTB~109, 1E~1841$-$045 in Kes~73,
and SGR 0526$-$66 in the LMC SNR N49 (Gaensler \etal\ 2001).
To these we may tentatively add \cco\ in \ctb, and
1E~1547.0$-$5408 because of its possible radio shell G327.24$-$0.13
\citep{gel07}.  In addition to providing age estimates,
such associations are
valuable for inferring properties
such as natal kick velocity,
and energetics of a supernova that is
able to create a magnetar (e.g., Vink \& Kuiper 2006).
The radio emitting AXP XTE~J1810$-$197 is the only
magnetar that has a well measured proper motion,
corresponding to $v_t = 212 \pm 35\ d_{3.5}$ km~s$^{-1}$ \citep{hel07},
which is not unusual compared to ordinary young neutron stars.
\citet{aha08a} have already pointed out that the
displacement of \cco\ from the center of the shell
of \ctb\ would require a transverse velocity of
$\sim 1000$ km~s$^{-1}$ to the east for an age of 5000~yr.
The velocity would be even larger if an association with
the historical supernova of 373 AD is assumed.
However, \citet{aha08a} also note that the brighter X-ray and radio
emission on the eastern side (see Figure~\ref{fig1}) could
indicate a higher density density there, implying that
the pulsar is closer to the explosion center, reducing
its inferred velocity.

In the case of \psr, the lack of secure spin measurements
means that the alternative of a low magnetic field neutron
star is not yet ruled out.  Its tentative 1~s period is
intermediate between those of the magnetars (2--12~s),
and the central compact objects (CCOs) in SNRs
that manifestly have weak fields and spin periods between
0.1 and 0.4~s \citep{got08}.  CCOs are not
clearly distinguished from AXPs on the basis of their X-ray
spectra alone, in the absence of spin-down data.
The temperature(s) and X-ray luminosity
required to fit the spectrum of \psr\ are similar to those of
CCOs, which require either one or two blackbodies
of $kT$ between 0.2 and 0.6~keV, and $10^{33} \le L_x \le 10^{34}$
erg~s$^{-1}$.  However, a fraction of
magnetars are observed for many years in
quiescent states of $10^{33} \le L_x \le 10^{34}$,
with spectra that are similar to CCOs. Examples
are the transient AXPs XTE J1810$-$197
\citep{got07}, 1E~1547.0$-$5408 \citep{gel07}, and
CXO J164710.2$-$455216 \citep{mun06}, and
SGR 1636$-$41 \citep{mer06}.  Because such low-luminosity
sources are difficult to classify before
they are seen in outburst, they may represent the majority
of magnetars.  Eventually, magnetars show X-ray
variability, whereas weak $B$-field CCO pulsars have not.
The classification of \psr\ remains uncertain until
either X-ray variability is established or its spin parameters
are measured.

\section{Conclusions}

We have identified two candidate magnetars in SNRs
based on their X-ray spectra, probable variability, and in
one case, \cco\ in \ctb, a typical magnetar pulse period
of $P=3.82$~s.  We have a second timing observation of each
candidate planned that will measure
its all-important period derivative, from which the spin-down
luminosity, characteristic age, and dipole magnetic field strength
will be derived.  This will test the magnetar hypothesis for
\cco.  Evidence of a hard spectral excess also needs
to be confirmed and explored at higher energy.

In the case of \psr\ in \snr, only marginal
evidence of a 1.01~s period is found.
However, given its soft X-ray spectrum typical of magnetars,
a period such as this is expected and is of great interest.
It lies between the low end for magnetars
(1E~1547.0$-$5408; $P=2$~s) and the transitional object PSR~J1846$-$0258
($P=0.32$~s).  With such a short period, it is
possible that the $\dot E$ of \psr\ exceeds the TeV luminosity
of \tev, $\approx 4.0 \times 10^{34}$ erg~cm$^{-2}$ s$^{-1}$,
or did so in the recent past.
Even if this candidate rotation period is not confirmed, it is
possible that the true period will be discovered in more sensitive
observations of \psr\ that will establish it as a magnetar.
On the other hand, if further spin measurements show that \psr\ is
a low $B$-field CCO, then its spin-down power was never
sufficient to power \tev.  In that case, the SNR \snr\ must
be responsible for the TeV emission.  Determining the spin
properties of \psr\ is therefore a crucial step in testing
whether old SNRs can power TeV sources.

\acknowledgements

We thank Elizabeth Galle and Nicholas Lee at the \chandra\ Helpdesk
for quickly resolving issues with the simulation of the point source response.
Support for this work was provided by the National Aeronautics and Space
Administration through \chandra\ Award Number GO9-0063X
issued by the \chandra\ X-ray Observatory Center, which is operated by the
Smithsonian Astrophysical Observatory for and on behalf of NASA under
contract NAS8-03060.
This investigation also used observations obtained with \xmm, an ESA
science mission with instruments and contributions directly funded by
ESA Member States and NASA.

\end{document}